\documentclass[useAMS,usenatbib,usegraphicx,fleqn]{mn2e}

\usepackage{color}

\usepackage{amsmath}
\usepackage{times}

\DeclareMathAlphabet{\mathpzc}{OT1}{pzc}{m}{it}

\newcommand\kms{\ensuremath{{\rm \, km\, s}^{-1} }}

\newcommand\zs{z_{\rm s}}
\newcommand\zsb{\bar z_{\rm s}}

\newcommand\Ds{D_{\rm s}}
\newcommand\Dos{D_{\rm 1s}}
\newcommand\Dts{D_{\rm 2s}}

\renewcommand\le\oldleq
\renewcommand\ge\oldgeq

\def\beq{\begin{equation}}
\def\eeq{\end{equation}}

\def\apj{ApJ}%
\def\apjl{ApJ}%
\def\aap{A\&A}%
\def\mnras{MNRAS}%
\title[Three-dimensional Microlensing]
      {Three-dimensional Microlensing}
\author[Mao, Witt \& An]
{Shude Mao$^{1,2}$, Hans J. Witt$^{3}$ and Jin H. An$^{1}$
\\$^1$ National Astronomical Observatories, 20A Datun Road, Chinese Academy of
Sciences, Beijing, 100012, China
\\$^2$ Jodrell Bank Centre for Astrophysics, The University of
Manchester, Alan Turing Building, Manchester M13 9PL, UK
\\$^3$ Im Hollergrund 76, 28357 Bremen, Germany
}
\begin{document}
\include{journaldefs}
\date{Accepted ...... Received ...... ; in original form......   }

\pagerange{\pageref{firstpage}--\pageref{lastpage}} \pubyear{2013}
\maketitle
\label{firstpage}

\begin{abstract}
We study three-dimensional microlensing where two lenses are located at
different distances along the line of sight. We formulate the lens equation
in complex notations and recover several previous results. There are in total either 4 or 6 images, with an equal
number of images with positive and negative parities. We find that the
sum of signed magnifications for six image configurations is unity. Furthermore,
we show that the light curves can be qualitatively different from those for binary
lensing in a single plane. In particular, the magnifications between a `U'-shaped caustic crossing can
be close to unity, rather than having a minimum magnification of 3 as 
in the single plane binary lensing. There is only a small probability 
three-dimensional microlensing events will be seen in microlensing toward the 
Galactic centre. It is more likely they will be first seen in cosmological microlensing.
\end{abstract}
\maketitle


\section{Introduction}
\label{sec:introduction}

Most gravitational lensing studies assume the thin lens approximation
where all the mass distribution is collapsed into a single plane. 
\cite{erdl93} first provided a complete classification of
the critical curves and caustics for two point lenses at different distances. 
The critical curves, caustics and bounds on the number of images of $N$
point lens in multiple planes have been further studied
using Morse theory \citep*{levi93, pett95a, pett95b, pett97, pett95w}.
The magnification probability distribution due to lenses in
multiple planes has also been explored by \cite{pei93} and \cite{lee97}.

For lensing by galaxies (rather than by point lenses such as stars),
\cite{koch88} first studied two-plane lensing, and concluded that 10\%
of gravitational lenses may be due to
multiple-plane lensing. This is roughly consistent with the CLASS
survey statistics within which one lens out of $\sim 20$ was found to be due to two galaxies at different redshifts \citep*{brow03, chae01}.
\cite*{wern08} studied the number of rings formed by
lenses in two different planes, concluding there are at maximum three rings in general.
Models have also been proposed for specific lenses, for example, \cite{chae01}
studied the case for B2114+022 \citep{Augu01} while \cite{Miho01} presents a two-plane lens model
for the system Q2237+0305. 

These previous studies concentrated on image numbers,
critical curves and caustics except \cite{pei93} and \cite{lee97}.
In this work, we shall study the light
curves due to three-dimensional microlensing. We first recast the lens
equations into complex form, and found new derivations in terms
of caustics and critical curves. We then
study the light curves in two point lens microlensing, find some
qualitatively different features in the light curve (see also \citealt{lee97})
and provide some analytic insights into the behaviour.

The structure of the paper is as follows: in \S\ref{sec:theory}, we
outline the theory of three-dimensional microlensing, in \S\ref{sec:results} we show
examples of light curves, and in \S\ref{sec:discussion} we then discuss
how such effects may be observable in Galactic and cosmological microlensing.

\section{Theory}
\label{sec:theory}

The basic formalism of three-dimensional microlensing for two-point
lenses are spelt out by \cite{erdl93}. Following their approach,
we align the optical axis with the line from the observer to the nearest
point lens, that is, the first lens is at the origin $(0,0)$.
The first lens has mass $M_1$ and is at distance $D_1$;
the second, farther-away lens has mass $M_2$ at distance $D_2$ ($D_1 \le D_2$). 
In complex notations, the lens equation can be written as
\beq
w = z - {\beta m_1 \over \bar{z}}, ~~
\zs = z - {m_1 \over \bar{z}} - {m_2 \over \bar{w} - \bar{z}_2},
\label{eq:lens1}
\eeq
where again the first lens is set at the origin,
$z_2$ is the second lens position
(in principle, we can put the second lens on the $x$-axis
without any loss of generality and then $z_2=\bar{z}_2$),
$w$ is where the light ray lands on the second lens plane,
$\zs$ is the source position, $m_1$ and $m_2$ are scaled
lens masses with
\beq
m_1 \equiv {M_1 \over M_1 + \alpha M_2}, ~~ m_2 \equiv 1 - m_1,
\eeq
and
\beq
\alpha \equiv {D_{1} \Dts \over \Dos D_2}, ~~
\beta \equiv {D_{12} \Ds \over \Dos D_2}.
\eeq
The lengths are normalised to the Einstein radius 
\beq
r_{\rm E} = \sqrt{ {4 G \Dos D_1 \over c^2} (M_1 + \alpha M_2)},
\label{eq:rE}
\eeq
and all the distances have their usual definitions, e.g.,
$\Dos$ is the distance from the first lens to the source,  $D_1$ is
the distance to the first lens, and $\Ds$ is the distance to the source.

Notice that the parameter $\beta$ specifies how three-dimensional the system
is. If the two lenses are in the same plane, then $\beta=0 (\alpha=1)$,
and the Einstein radius is simply that corresponding to the total mass
$M_1+M_2$. On the other hand, if $D_2
\rightarrow \Ds$, $\alpha \rightarrow 0$, $\beta \rightarrow 1$ and the
Einstein radius defined in eq. (\ref{eq:rE}) is determined by $M_1$ only.
In general, $0\le \beta \le 1$.

In this paper we shall adopt Euclidean geometry
which is applicable to microlensing in the Galaxy. In this case, 
\beq \label{d1d2}
\beta = { (d_2 - d_1 ) \over (1- d_1) d_2 }, ~~ \alpha = 1 - \beta,
\eeq
where we define
\beq
d_1 \equiv {D_1 \over D_s}, ~~ d_2 \equiv {D_2 \over D_s}.
\eeq
All our results can be trivially generalised to cosmological microlensing
by adopting angular diameter distances, although $\alpha=1-\beta$ is no
longer true. 

\subsection{Image positions}

Eliminating $\bar w$ from eq.~(\ref{eq:lens1})
by means of inserting the complex conjugate of the first to the second results in
\beq \label{eq:leqs}
\zs = z - {m_1 \over \bar{z}} - {m_2 z \over (\bar{z}-\bar{z}_2) z - \beta m_1}.
\eeq
The image positions are thus found to be zeros of the polynomial,
\beq \label{eq:pol}
f(z,\bar z)=(z\bar z-\zs\bar z-m_1)(z\bar z-\bar z_2z-\beta m_1)-m_2z\bar z.
\eeq
Considered as independent variables,
$(z,\bar z)$ is then the simultaneous solution of two equations
$f(z,\bar z)=\overline{f(z,\bar z)}=0$.
For the polynomial in eq.~(\ref{eq:pol}),
the Bernstein theorem \citep[see also \citealt{polynom}]{Be75}
indicates that there in general exist six non-zero complex solutions
provided that none of $\zs,z_2,\beta,m_1,m_2$
is zero (there are five solutions if $\beta=0$
whilst the number of non-zero solutions is four if either $\zs=0$ or $z_2=0$).
This is also verified more prosaically by further eliminating $\bar z$ from
eq.~(\ref{eq:leqs}) using its complex conjugate.
In principle we can achieve this by computing the resultant of
$f(z,\bar z)$ and $\overline{f(z,\bar z)}$ considered as 
the polynomials of $\bar z$, which is easily done using any modern
symbolic computer algebra software, e.g.,
{\sc Maple}\textsuperscript{\texttrademark} or
{\sc Mathematica}\textsuperscript{\textregistered}.
We then obtain the sextic polynomial of $z$,
whose coefficients are given in Appendix A.
Therefore the number of images for the lens system here
is bounded above by six, which is in fact achieved when
a source is inside caustics.

The lens system under consideration is localised,
and so the images for $\zs\to\infty$ are obtained
either for divergent bending angles,
which formally occur when the light ray falls exactly on the lens,
or for $z\to\infty$. Counting two solutions of $w=z_2$
as well as $z=0$ and $z=\infty$, we then find that there must be
four images as $\zs\to\infty$, which is the number of
images for a source outside caustics. We also find that
the images corresponding to $z\to\infty$ and $z\to0$ respectively have
the positive and the negative parity, whereas the parities of the two images
for $w=z_2$ are opposite to each other. Since the index
theorem for image parities still holds, a pair of images having opposite parities
form or merge when the source crosses caustics.
Hence, we have always an equal number of positive
and negative parity images \citep[see][]{erdl93}.

\subsection{Magnifications}

Using external products,
it is trivial to generalise the expression in \cite{witt90}
to the three-dimensional case to obtain (see \citealt{rhie97})
\begin{gather}
\det J=
\upartial\zs \bar{\upartial}\zsb - \upartial\zsb \bar{\upartial}\zs;
\\
\upartial\zs = 1 + {\beta m_1 m_2 \over h^2}, ~~
\bar{\upartial}\zs = {m_1 \over \bar{z}^2} + {m_2 z^2 \over h^2}.
\end{gather}
where $h\equiv z (\bar{z} - \bar{z}_2) - \beta m_1$
whilst $\upartial\zs\equiv\partial\zs/\partial z$ and
$\bar{\upartial}\zs\equiv\partial\zs/\partial\bar z$ etc.\
(also note $\overline{\upartial\zs}=\bar{\upartial}\zsb$ and
$\overline{\upartial\zsb}=\bar{\upartial}\zs$).
The magnification of the image at $z$ is then given by $\mu=(\det J)^{-1}$.

The critical curves are defined as the locus of $\det J=0$,
that is, $\upartial\zs\bar{\upartial}\zsb
=\upartial\zsb\bar{\upartial}\zs$, which is also
equivalent to $\lvert\upartial\zs\rvert=\lvert\upartial\zsb\rvert$
and $\lvert\upartial\zs/\upartial\zsb\rvert=1$.
We may then write a parametric representation
of the critical curves
\begin{equation}\label{eq:critp}
{\rm e}^{2{\rm i}\varphi}{\upartial\zs}={\bar{\upartial}\zs}, ~~
{\rm e}^{-2{\rm i}\varphi}{\bar{\upartial}\zsb}={\upartial\zsb}.
\end{equation}
We note the linearised lens mapping near the point on critical curves
follows as
\begin{equation}
\delta\zs(\delta z)
= \upartial\zs\,\delta{z} + \bar{\upartial}\zs\,\delta\bar{z}
= \left( 1 +
{\rm e}^{2{\rm i}\varphi} {\delta\bar{z} \over \delta{z}} \right)
\upartial\zs\,\delta{z},
\end{equation}
and thus we find that $\delta\zs({\rm i}{\rm e}^{{\rm i}\varphi})=0$
-- that is, ${\rm i}{\rm e}^{{\rm i}\varphi}={\rm e}^{{\rm i}(\varphi+\upi/2)}$
corresponds to the critical direction on the image plane.
Note also that $\delta\zs({\rm e}^{{\rm i}\varphi})=
2{\rm e}^{{\rm i}\varphi}\upartial\zs$, but here $\upartial\zs$ is
not necessarily real. Hence the projection direction, ${\rm e}^{{\rm i}\varphi}$
mapped onto the source plane becomes rotated by $\xi=\arg(\upartial\zs)$
whilst ${\rm e}^{{\rm i}(\varphi+\xi)}$ is the tangential direction
to the caustics. In fact, we may alternatively parameterise the critical
curves through the tangential direction angle, $\chi=\varphi+\xi$,
using $\upartial\zs/\upartial\zsb=
(\upartial\zs)/({\rm e}^{-2{\rm i}\varphi}\overline{\upartial\zs})=
{\rm e}^{2{\rm i}\chi}$.

Two independent equations of $z$ and $\bar{z}$ in eq.~(\ref{eq:critp})
may be transformed again into a polynomial
of $p[z({\rm e}^{2{\rm i}\varphi})]$, which can be solved
in principle for each $\varphi \in [0,\upi[$. However, unlike
the single-plane case, each equation in eq.~(\ref{eq:critp})
involves both $z$ and $\bar{z}$. Once $\bar{z}$ is eliminated,
the polynomial $p(z)$ for a generic case is of 20th degree,
zeroes of which includes many spurious solutions which need
to be checked with the original equation of eq.~(\ref{eq:critp}).
In fact, the equation for the critical curves can be solved
more straightforwardly in polar coordinates \citep[see also][]{erdl93}
by setting $z=\lvert z\rvert{\rm e}^{{\rm i}\theta}$ in
$\lvert\upartial\zs\rvert^2=\lvert\upartial\zsb\rvert^2$,
which reduces to a quadratic equation on $\cos\theta$ at a fixed $\lvert z\rvert$.


\subsection{Magnification sum invariants}

For binary lensing in a single plane, \citet{witt94} and \citet{rhie97}
showed that when a source is inside a caustic (five-image configurations),
the sum of signed magnifications $\sigma=\sum_i\mu_i$ is always unity.
Based on the method developed by
\citet[][see also \citealt{dalal01,evan02}]{evan01},
\citet{AE06} extended this to show that, if the lens equation is
given by $\zs=z-\overline{s(z)}$ with an analytic function $s(z)$,
then the sum of signed magnifications $\sigma$ is related to
the coefficient of the Laurent series expansion of $[g(z)]^{-1}$
as $z\to\infty$ such that $[g(z)]^{-1}\simeq\sigma z^{-1}+\mathcal O(z^{-2})$.
Here $g(z)=z-\zs-\bar s[\zsb+s(z)]$ and $\bar s(w)=\overline{s(\bar w)}$.
For the case that $\lvert\lim_{z\to\infty}s(z)\rvert<\infty$,
this then implies $g(z)\sim z$ for a generic source position $\zs$,
and it follows that $\sigma=1$.

As the deflection term of eq.~(\ref{eq:leqs}) contains $z$ as well as $\bar z$,
their result is not applicable here. However, \citet{wern07}
has shown that if the lens equation is given by $\zs=z-s(z,\bar z)$
where $s(z,\bar z)$ is a bivariate rational function of $(z,\bar z)$
such that the degree of the polynomial in the denominator is greater
than that in the numerator (i.e.\ $s\to0$ as $z\to\infty$),
then the holomorphic Lefschetz fixed-point formula \citep[see][]{alg-geo}
from algebraic geometry and topology applies,
and subsequently implies that $\sigma=1$. It is easy to find that
eq.~(\ref{eq:leqs}) indeed meets the condition,
and thus the result of \citet{wern07} applies here. That is to say,
the sum of signed magnifications for six-image configurations
should be exactly unity
\beq
\sum_{i=1}^6 \mu_i = 1,
\label{eq:signed_mu}
\eeq
which we have also verified numerically.

\section{Example light curves}
\label{sec:results}

For illustration, we study a scenario where the two lenses have
$M_1=0.7$ and $M_2=0.3$ (the values of $m_1$ and $m_2$ will vary as
$\beta$ changes). We set the source distance at 1, and put the first
lens is at distance $d_1=0.7$. The distance to the second lens ($d_2$) is varied
to see the differences in the light curves. Both lenses
are assumed to be static, with the first lens at the origin and the
second lens at $z_2=(0.8, 0.0)$. The source trajectory is horizontal,
with the vertical coordinate being 0.03.

\begin{figure*}
\begin{centering}
\includegraphics[width=.5\hsize]{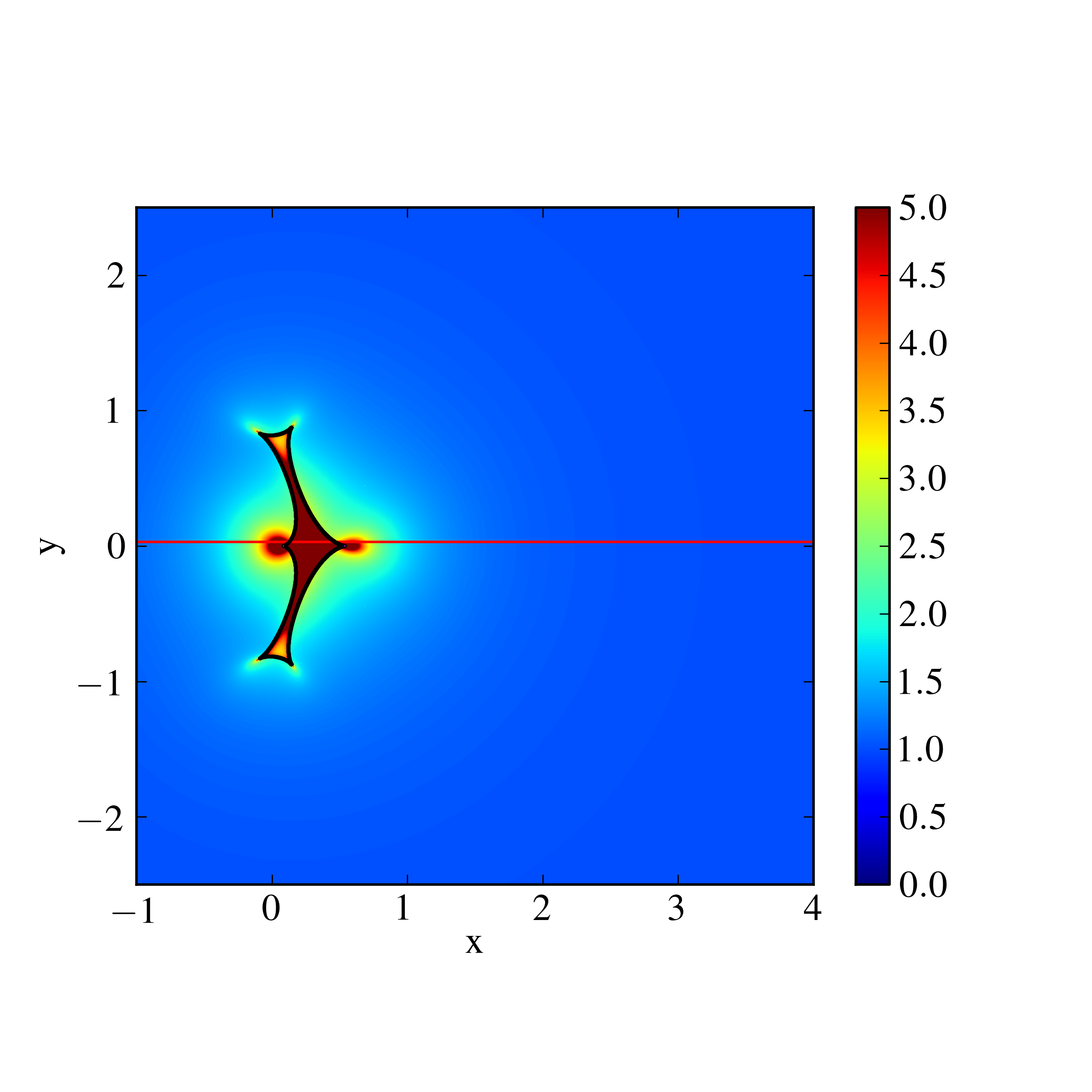}
\includegraphics[width=.47\hsize]{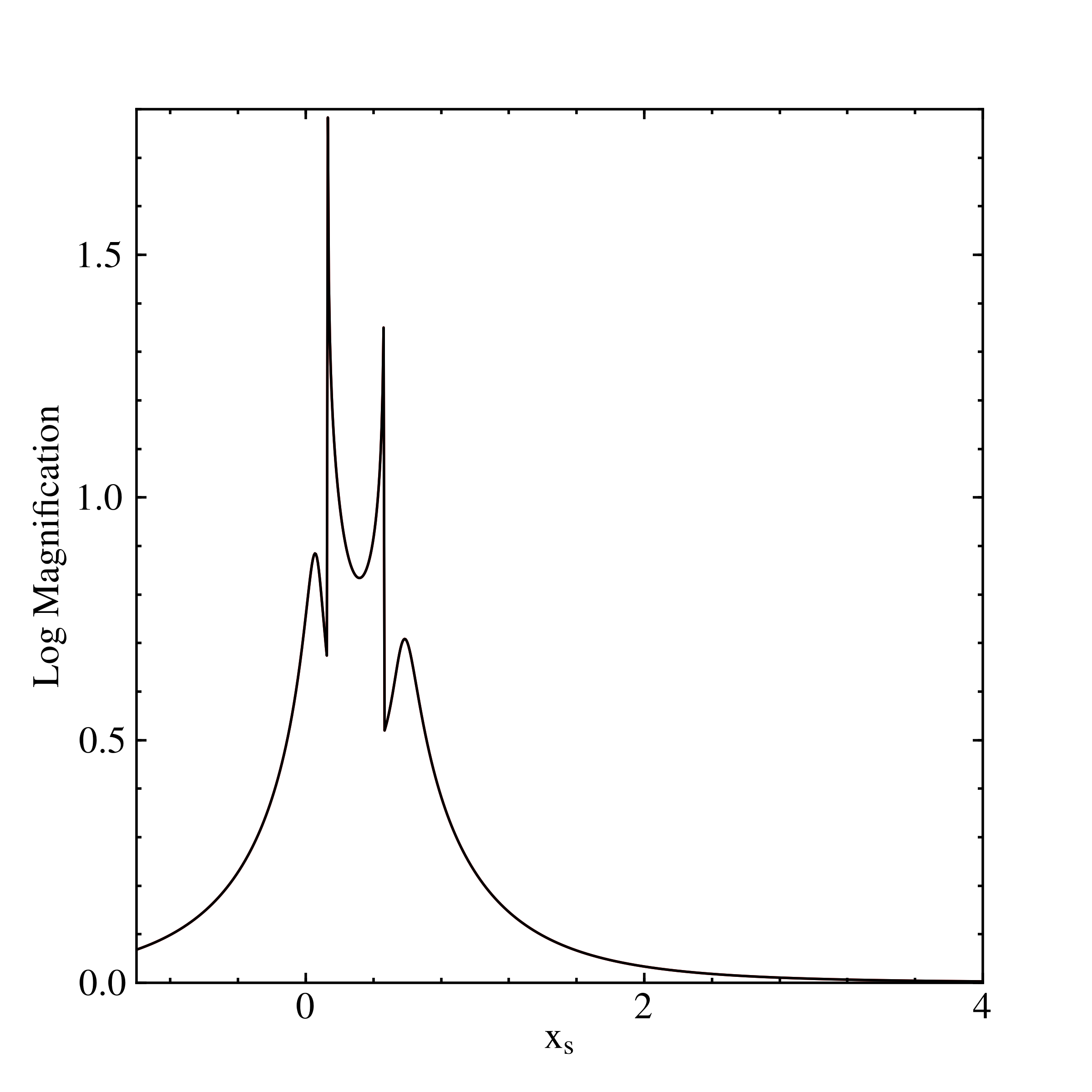}
\end{centering}
\caption{Left: Example caustics, source trajectory and magnification
  patterns for $d_2=0.7$. The two lenses have masses $M_1=0.7$ and
  $M_2=0.3$ with the first lens at the origin and the second at $(0.8, 0.0)$.
  Right: Light curve for the horizontal source trajectory with $y_s=0.03$
  shown in the left.}
\label{fig:lc-0.7}
\end{figure*}

\begin{figure*}
\includegraphics[width=.5\hsize]{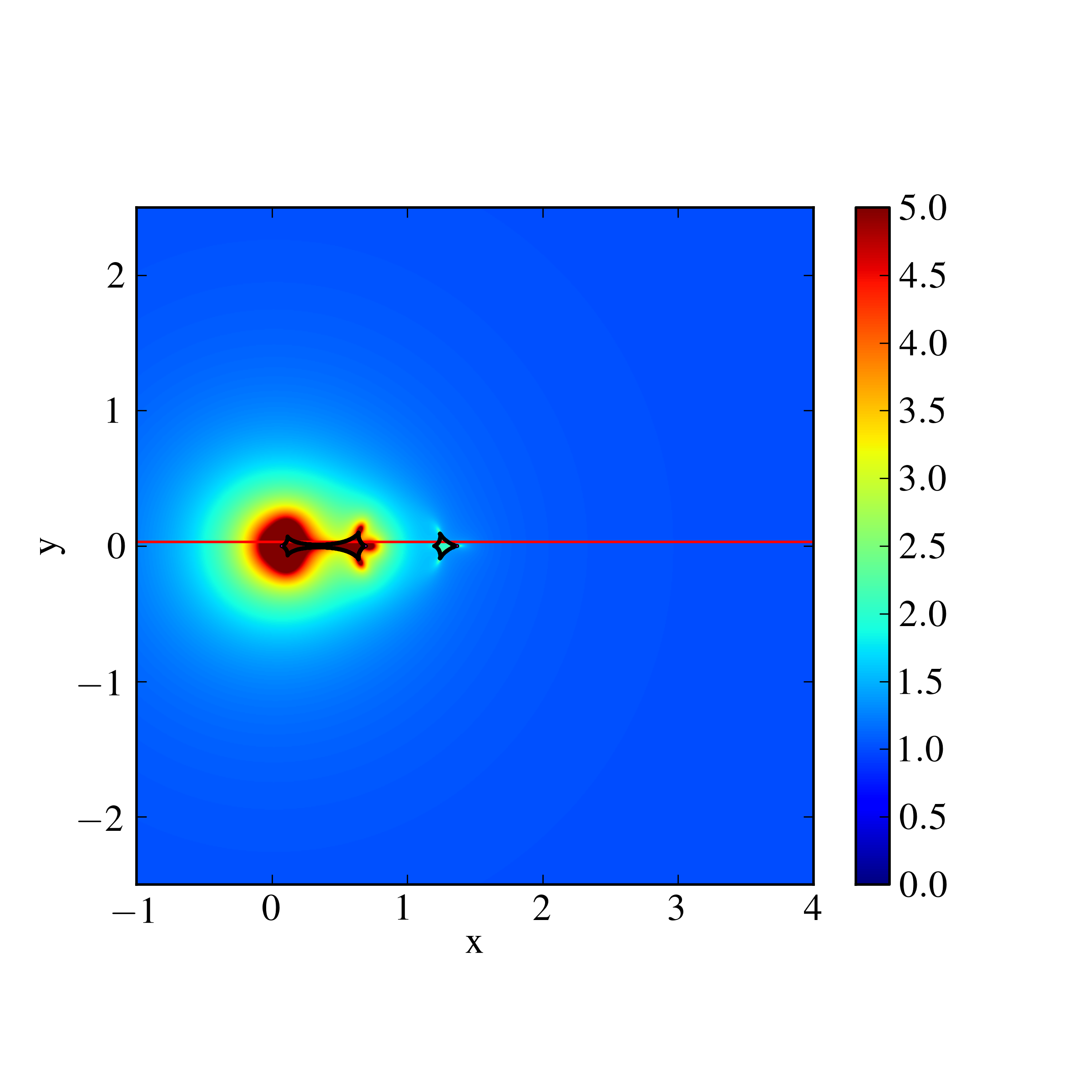}
\includegraphics[width=.47\hsize]{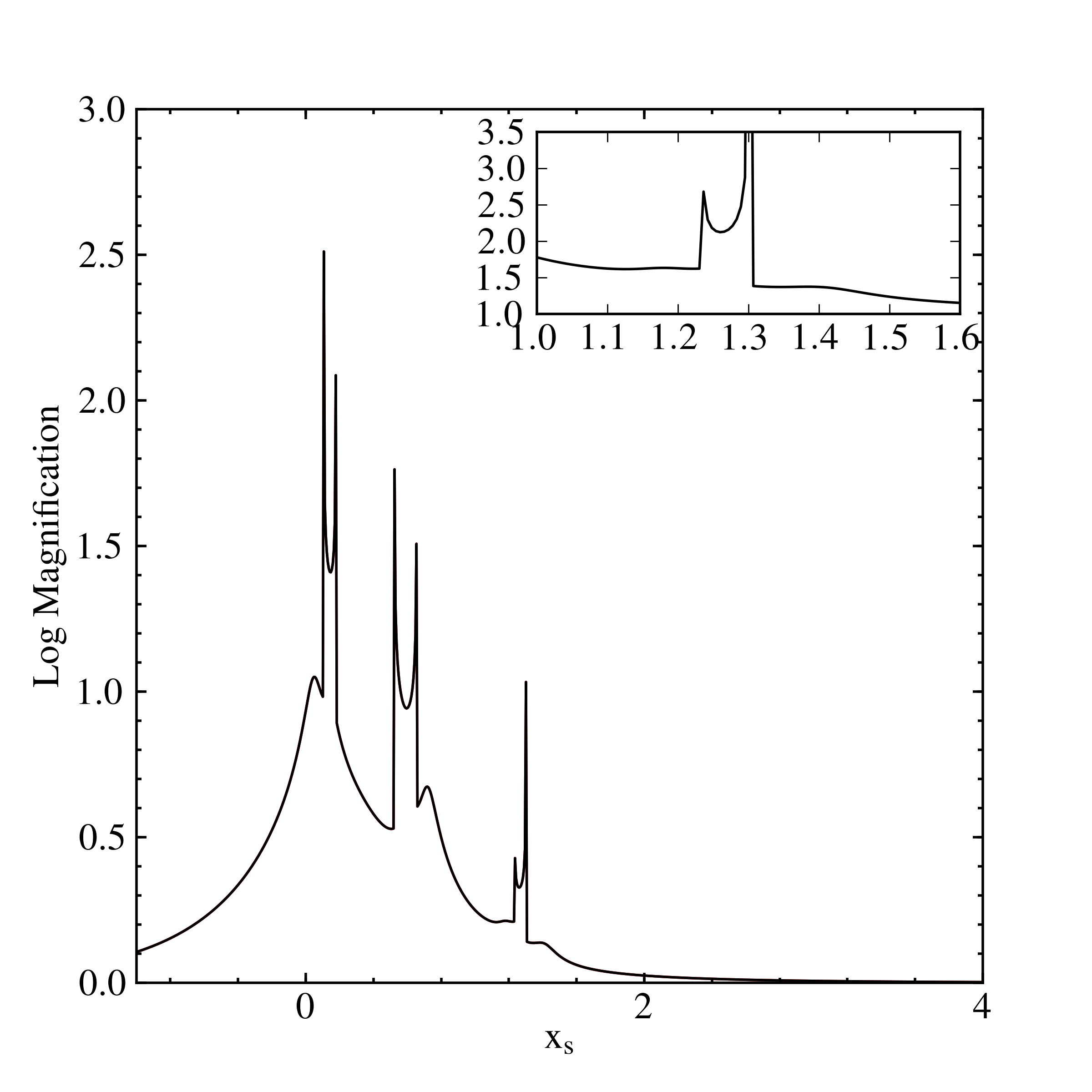}
\caption{Left: Example caustics, source trajectory and magnification
  patterns. Right: Light curve for the horizontal source trajectory shown in
  the left for $d_2=0.9$. The inset shows the zoomed in view of the second
  caustic crossing with magnification on linear scale.
  Notice that the magnification between the `U'-shaped caustic crossing
  is small, $\sim 2.3$, still smaller than the minimum magnification
  between caustic crossings for binary lenses in a single plane.}
\label{fig:lc-0.9}
\end{figure*}

In Figs. \ref{fig:lc-0.7} to \ref{fig:lc-0.9} we show the magnification
patterns and light curves for $d_2=0.7, 0.8$, and 0.9,
respectively. At $d_2=d_1=0.7$, we recover the single plane lensing, and 
there is only one joint caustic on the left. As $d_2$ increases,
the three-dimensionness of lensing increases, and an additional caustic on
the right appears. But this is a weak/faint caustic, as a source crosses it,
the magnification between the crossing can be very small. For
$d_2=0.8$, it is 1.15, much smaller than the minimum magnification 
of 3 for binary lensing in a single plane \citep{witt94}. This is
because the faint caustic (on the right) has been strongly deflected by the first lens $M_1$, and thus the magnification is lower. 
This behaviour is known in \cite{lee97}, although we recovered this without
knowing their results.

As $d_2$ further increases to 0.9, the additional, right-most caustic moves towards
the main caustic on the left, and the magnification between the
`U'-shaped caustics becomes higher, of the order of 2.3, still lower
than the minimum value of 3 as in the single-plane binary lensing.

In Appendix B, we quantitatively study the trend of the magnification
with $\beta$ for a source on the $x$-axis. We show how the behaviour seen in the figures can be understood.

\section{Summary and Discussion}
\label{sec:discussion}

In Galactic microlensing, the probability by microlensing of one star is
of the order of $\tau\sim 10^{-6}$, and so the probability of microlensing
by two independent stars along the
line of sight is $\tau_2 \approx \tau^2 \sim 10^{-12}$.
Each year, we monitor about $N_\star \sim 2\times 10^{8}$ stars,
and so the event rate for 3D microlensing is
$N_\star \times \tau_2/t_{\rm E}$,
where $t_{\rm E} \approx 0.1~\mbox{yr}$ is the typical microlensing event, so this
gives about $2\times 10^{-3}~\mbox{events yr$^{-1}$}$, which is quite low.
Notice that in principle the farther lens is also lensed by the closer
one and its light curve will be superimposed on that of the background source.
However, in practice, this effect may be small because the lenses are expected
to be quite faint.

In a single plane Galactic microlensing, the motion of the centre of
the mass of the binary can be absorbed into the motion of the source, although the binary rotation (of the order of
$\sim 10\kms$) can still be
observed (\citealt{albr00, an02, jaro05, hwan10, ryu10}; for predictions see, \citealt{pkm11, pmk11}).
For three-dimensional lensing,  the two lenses will move independently with velocities of $100\kms$, and so
the distance between the lenses, and as a result critical curves and
caustics, will change as a function of time more rapidly,
which may further diversify the light curves. Fig. \ref{fig:move} shows an example of a light curve due to a moving lens. The rate of change in the light curve will depend on the relative motions. In particular, motions parallel to the two lenses may have a bigger impact on the caustics than perpendicular motions. This can be understood analytically in terms of the motions of caustics (see Appendix B).

So far, we have focused on ``resonant'' microlensing in two planes, it should be mentioned that if the lenses are widely separated then they act as unrelated lenses, and the probability of observing such cases will be comparable to or even larger than the ``resonant'' microlensing, depending on the separation distributed of close/wide binaries. Indeed, repeated microlensing events due to binary lenses in a single plane have been predicted and observed (\citealt{dm96, ds99, skow09}). In this case, the two microlensing peaks will have different timescales but the same colour. They can be differentiated from single-plane binary source events which will have almost the same timescale (modified by the rotation of the binary) and (possibly) different colours \citep{GH92}. The wide-binary lens event in a single plane may be difficult to tell apart from wide three-dimensional microlensing events; the latter, as we argued above, may have much more different timescales than wide single-plane lensing events.

A more likely case for 3D microlensing is cosmological microlensing by stars located
in two lens galaxies at different redshifts.
We know multiple images can be produced by galaxies at different
redshifts, as already seen in the CLASS lens B2114+022 \citep{chae01}.
In this case, the optical depth for microlensing in each plane is of
the order of 0.1--1,
and so the two-plane microlensing has a high chance of occurrence.
It will be very interesting to explore how the magnification maps are
qualitatively different from that in a single microlensing case. Our
preliminary study shows that indeed faint caustic crossing events found in this
paper also occur, as also found by \cite{lee97}.
It will be interesting
to examine this further for specific cases since {\sl LSST}\footnote{www.lsst.org}
will discover
approximately $10^4$ new gravitational lenses while simultaneously
obtaining their light curves. Some of these ($\approx$ 10 per cent) will be
three-dimensional lenses. Notice also that the second lensing galaxy in
this case should in principle be ``microlensed'' by the first galaxy,
although in practice, due to the extended size nature of the galaxy, the
effect may be small unless there are bright compact sources such as
a new supernovae or an active galactic nucleus at the centre of the second galaxy. 

\section*{Acknowledgments}
We acknowledge the National Astronomical Observatories (NAOC) and
Chinese Academy of Sciences (CAS) for financial support (SM and JHA). 
SM acknowledges the Max-Planck Institute for Astrophysics in Garching
for travel support where this work was completed.
HJW thanks the NAOC and in particular the gravitational lensing and
galaxy group for hospitalities during a visit. 


\clearpage

\begin{figure*}
\includegraphics[width=.5\hsize]{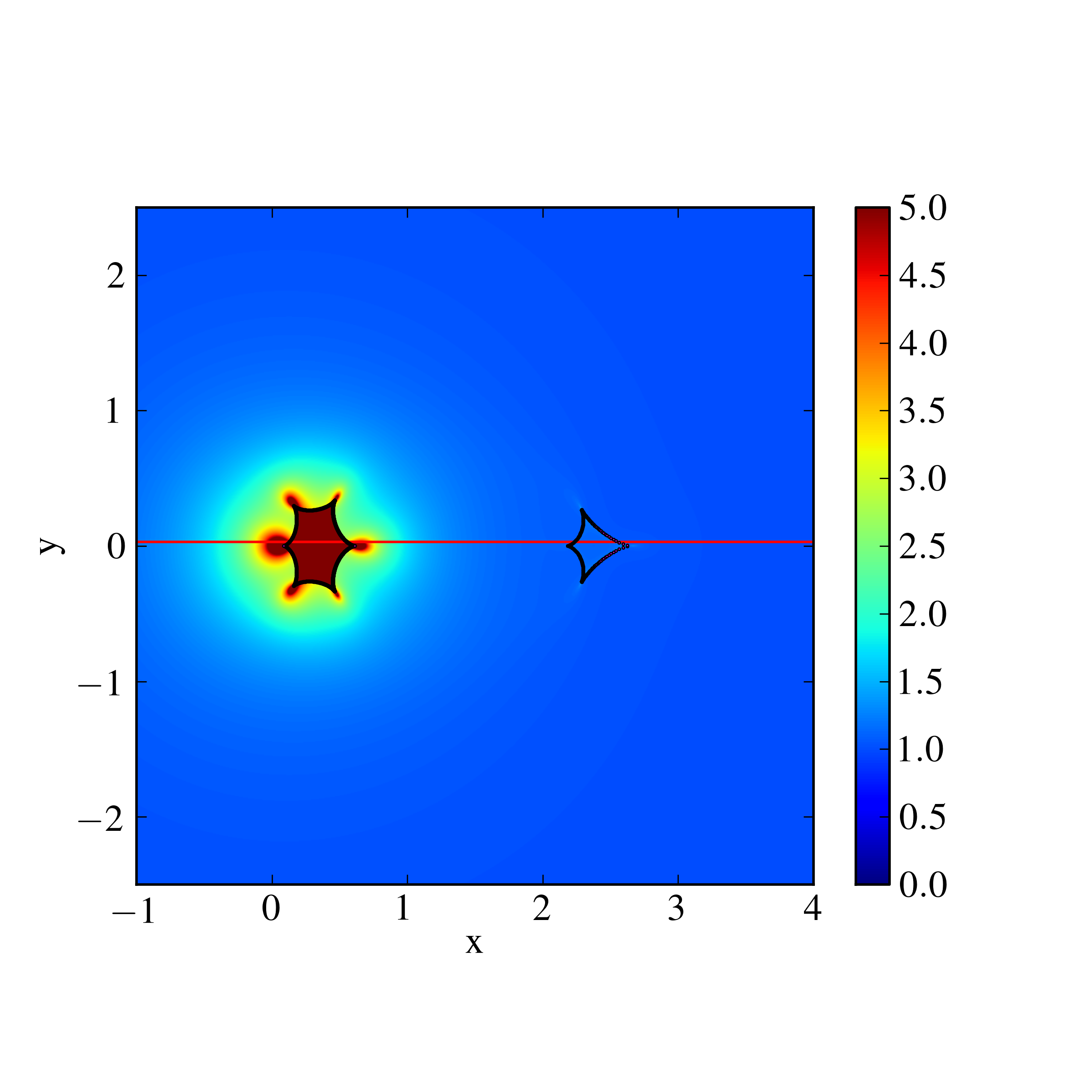}
\includegraphics[width=.47\hsize]{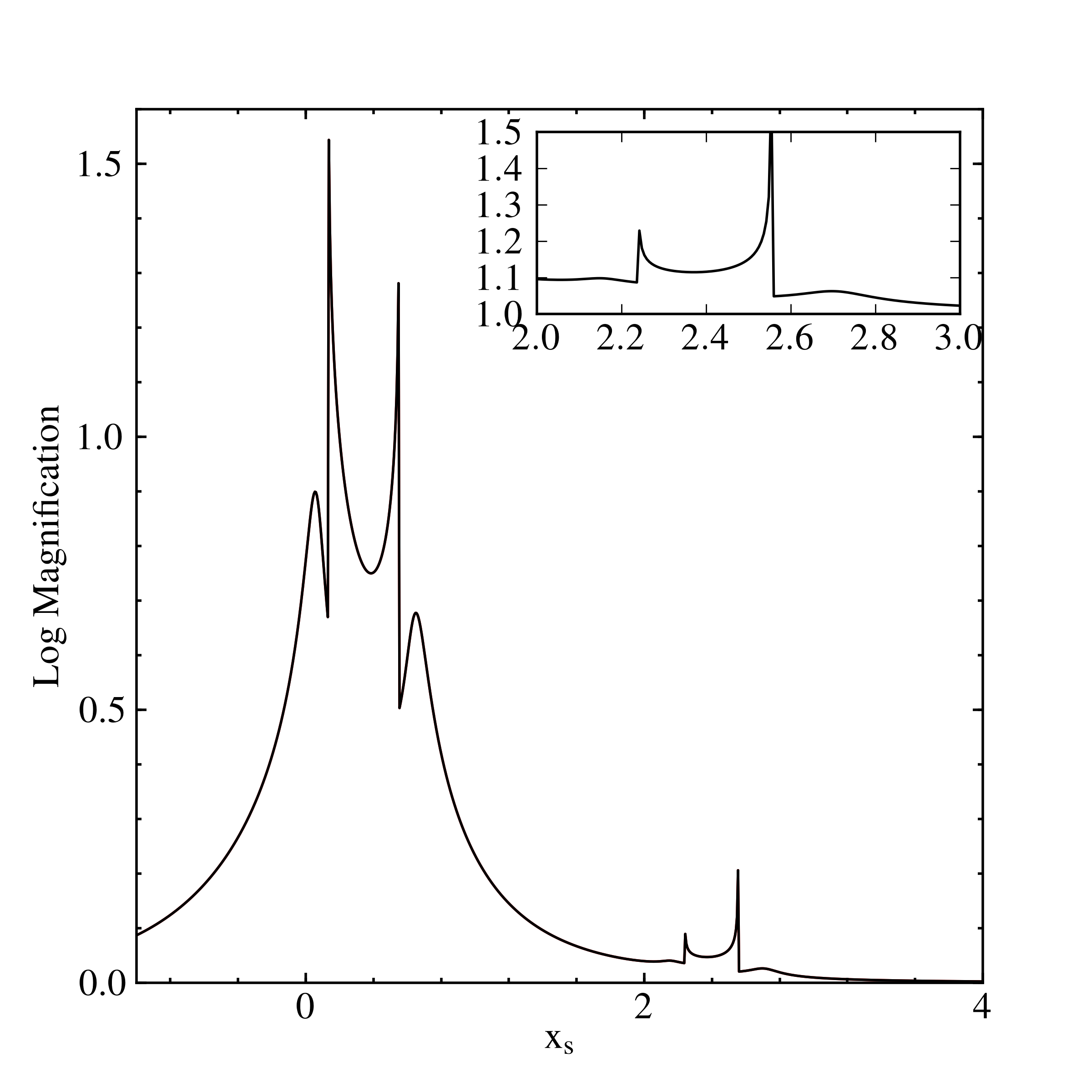}
\caption{Left: Example caustics, source trajectory and magnification
  patterns for $d_2=0.8$. Right: Light curve for the horizontal
  source trajectory shown in the left.
  The inset shows the zoomed in view of the second
  caustic crossing with magnification on linear scale.
  Notice that the magnification between the `U'-shaped
  caustic crossing is small, $\sim 1.15$.}
\label{fig:lc-0.8}
\end{figure*}

\begin{figure*}
\begin{centering}
\includegraphics[width=0.5\hsize]{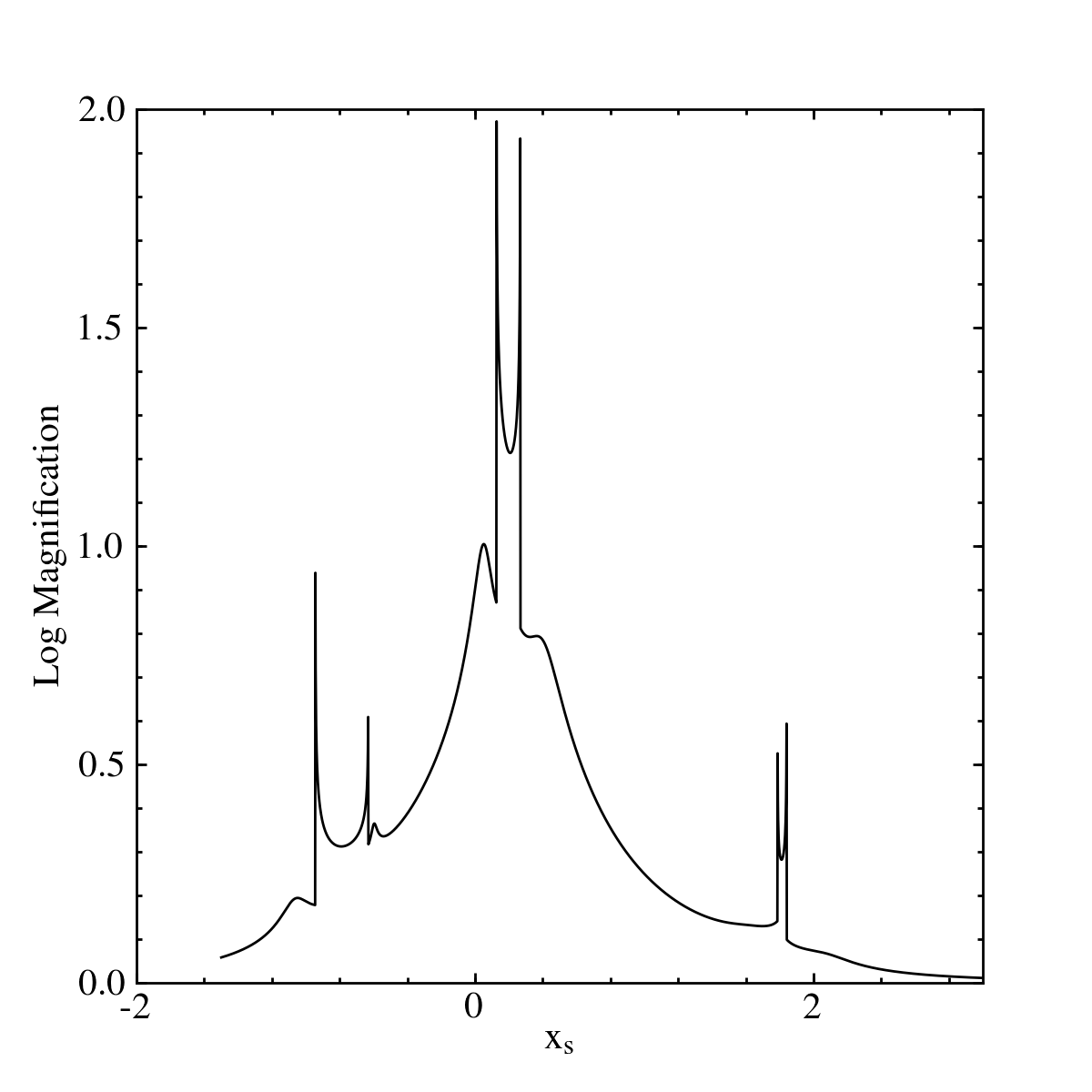}
\end{centering}
\caption{
Light curve for a horizontal source trajectory with $y_s=0.03$,
identical to the ones shown in Figs. \ref{fig:lc-0.7} to \ref{fig:lc-0.9}.
The two lenses have masses $M_1=0.7$ and
  $M_2=0.3$; the first lens is at a distance of $d_1=0.7$ and the second
lens at $d_2=0.9$ (the same as for Fig. \ref{fig:lc-0.9}).
The first lens is fixed at the origin (in projection) while the second lens is moving
with $dz_2/dt=0.54$. As the source moves from $-1.5$ to 3.5, the second lens
moves from $(-0.55, 0)$ to (2.15, 0).
}
\label{fig:move}
\end{figure*}

\clearpage
\appendix
\onecolumn
\section{Complex lens equation}

The coefficients for the resultant complex sextic polynomial
$g(z;\zs,\zsb)=\sum_{k=0}^6a_k z^k$ of the lens equation are given by
\begin{subequations}\begin{gather}
a_6=(1-\beta)(\zsb-\bar z_2)\bar z_2\zsb,
\\a_5=m_1(1-\beta)^2\bar z_2\zsb-m_2(\beta\zsb-\bar z_2)(\zsb-\bar z_2)
-\bigl[(2-\beta)z_2+(1-2\beta)\zs\bigr](\zsb-\bar z_2)\bar z_2\zsb,
\\\begin{split}
a_4=\bigl[2(1-\beta)z_2\zs+z_2^2-\beta\zs^2\bigr](\zsb-\bar z_2)\bar z_2\zsb
&+m_1(1-\beta)
\bigl[\bar z_2^2\zs-\beta z_2\zsb^2-2(1-\beta)(z_2+\zs)\bar z_2\zsb\bigr]
\\&-m_2\bigl[2(\zs+\beta z_2)\bar z_2\zsb-(z_2+\zs)(\bar z_2^2+\beta\zsb^2)\bigr],
\end{split}\\\begin{split}
a_3&=(\beta\zs-z_2)(\zsb-\bar z_2)z_2\bar z_2\zs\zsb
-m_1^2(1-\beta)^2(\bar z_2\zs+\beta z_2\zsb)
+m_2^2(\beta\zsb-\bar z_2)(\zs-z_2)
\\&\quad+m_1m_2\bigl[(1+\beta)(z_2\bar z_2+\beta\zs\zsb)
-2(\bar z_2\zs+\beta^2z_2\zsb)\bigr]
-m_2\bigl[(z_2\zsb+\bar z_2\zs)(z_2\bar z_2+\beta\zsb\zs)
-2(1+\beta)z_2\bar z_2\zs\zsb\bigr]
\\&\quad+m_1\Bigl\{
\bigl[(1-2\beta)z_2\zsb-(2-\beta)\bar z_2\zs\bigr](z_2\bar z_2+\beta\zs\zsb)
+\beta(z_2\zsb+\bar z_2\zs)^2+4(1-\beta)^2z_2\bar z_2\zs\zsb\Bigr\},
\end{split}\\\begin{split}
a_2=m_1\bigl[2(1-\beta)\bar z_2\zsb-\bar z_2^2+\beta\zsb^2\bigr]
(\beta\zs-z_2)z_2\zs
&-m_1^2(1-\beta)\bigl[
\beta(\bar z_2\zs^2-z_2^2\zsb)-2(1-\beta)(\bar z_2+\beta\zsb)z_2\zs\bigr]
\\&+m_1m_2\bigl[2(\bar z_2+\beta^2\zsb)z_2\zs
-(\bar z_2+\beta\zsb)(z_2^2+\beta\zs^2)
\bigr],\end{split}\\
a_1=m_1^3\beta(1-\beta)^2z_2\zs-m_1^2m_2\beta(\beta\zs-z_2)(\zs-z_2)
+m_1^2\bigl[(1-2\beta)\bar z_2+\beta(2-\beta)\zsb\bigr](\beta\zs-z_2)z_2\zs,
\\a_0=m_1^3\beta(1-\beta)(\beta\zs-z_2)z_2\zs.
\end{gather}\end{subequations}

The degree of the polynomial may be reduced if $a_0=0$ or $a_6=0$.
The case $\beta=0$, which corresponds to the single plane lensing,
results in $a_0=0$ and thus $z$ factors out of
the polynomial leaving a quintic quotient. Since $z=0$ is not a solution
of the lens equation, the maximum number of images of a single plane
binary lensing is five. If $\beta=1$ on the other hand, then both $a_0=a_6=0$
and so the polynomial becomes a quartic times $z$, leaving the maximum
four images. Similar reductions of the polynomial to a quartic are also
possible for $z_2=0$ (i.e.\ two lenses being aligned)
or the special source positions corresponding to
$\zs=0$, $\zs=z_2$ or $\zs=\beta^{-1}z_2$. However they can be
understood as particular cases of $\zs=\zeta z_2$ with $\zeta\in\mathbf R$,
which shall be discussed next.

In principle, without any loss of generality, the real axis can be chosen
such that the complex position of second deflector $z_2$ is positive real
(i.e.\ $z_2=\bar z_2>0$).
The case that both the source and the second deflector are
located on the real axis on the other hand represents the physical scenario
that the observer, two deflectors, and the source are all co-planar.
A geometric argument concerning this case  indicates that
there must be four images along the real axis as well.
These are found with eqs.~(\ref{eq:leqs}) or ~(\ref{eq:pol})
by setting $z=\bar z$ and $z_2=\bar z_2$,
which results in a quartic polynomial equation $g_4(z)=0$ where
\begin{equation}\begin{split}
g_4(z)&=z^4 - (z_2+\zs) z^3 + \bigl[ z_2 \zs - (\beta+1)m_1-m_2 \bigr]z^2
+ m_1 (z_2+\beta\zs) z + \beta m_1^2
\\&=(z^2-\zs z-m_1-m_2)(z^2-z_2z-\beta m_1)-m_2(z_2z+\beta m_1).
\end{split}\end{equation}
Given real $\zs,z_2\in\mathbf R$ and positive $\beta,m_1,m_2>0$, the equation
$g_4(z)=0$ possesses four real solutions $z\in\mathbf R$, which is shown
as follows. Let us suppose that $z_\pm$ are two zeroes of
$z^2-z_2z-\beta m_1$ (i.e.\ corresponding to $w=z_2$), namely,
\begin{subequations}
\begin{equation}
z_\pm=
\frac{z_2\pm\sqrt{z_2^2+4\beta m_1}}2
=\pm\frac{z_2}2
\left[\left(1+\frac{4\beta m_1}{z_2^2}\right)^{1/2}\pm1\right],
\end{equation}
and thus $z_-<0<z_+$ for $z_2>0$ (the following is still valid
for $z_2<0$ with $z_+\leftrightarrow z_-$).
Next $g_4(0)=\beta m_1^2>0$ whilst
\begin{equation}
g_4(z_\pm)=-m_2(z_2z_\pm+\beta m_1)
=-\left[\left(1+\frac{2\beta m_1}{z_2^2}\right)
\pm\left(1+\frac{4\beta m_1}{z_2^2}\right)^{1/2}\right]\frac{m_2z_2^2}2<0.
\end{equation}
\end{subequations}
Since $g_4(z)$ is a monic quartic, the intermediate value theorem
together the fundamental theorem of algebra indicates that $g_4(z)=0$
has four real solutions, -- one each in the intervals,
$]{-\infty},z_-[$, $]z_-,0[$, $]0,z_+[$, and $]z_+,\infty[$ --
all of which corresponds to true image positions.
The remaining two off-axis solutions, if any, must still be the zero
of the sextic $g(z;\zs,\zsb)$ with $\zs=\zsb$ and $z_2=\bar z_2$,
which is in fact divisible by $g_4(z)$, that is to say,
$g(z;\zs,\zs)=g_4(z)g_2(z)$ where
\begin{equation}
g_2(z)= (1-\beta) (\zs - z_2) \zs z_2 z^2
+ \bigl[ m_1 (1-\beta)^2 \zs z_2
+ (z_2 \zs - m_2) (\beta \zs - z_2) (\zs - z_2) \bigr] z
+ m_1 (1-\beta) (\beta \zs -z_2) \zs z_2.
\end{equation}
Since $\zs,z_2\in\mathbf R$,
the coefficients of $g_2(z)$ are all real, and
its two zeroes are either both real or a pair of complex conjugates.
Next, we find $f(z_a,z_b)=0$ where $z_a$ and $z_b$ are
the pair of zeroes of $g_2(z)$ and $f$ is the polynomial in eq.~(\ref{eq:pol}).
In other words, the complex conjugate roots of $g_2(z)=0$
are true off-axis images whereas its non-degenerate real zeroes
are spurious solutions.
Consequently, along the real axis, the source lies inside, on, and outside
caustics if the discriminant of $g_2(z)$ is negative, zero, and
positive respectively.
Some results for the corresponding magnifications
are explored in the next section.

If $z_2=0$ (that is, two lenses are aligned),
\begin{equation}
g_4(z)=z^4 - \zs z^3 - {\cal M} z^2 + \beta m_1 \zs z + \beta m_1^2,
\quad
g_2(z)=-\beta m_2\zs^2z
\end{equation}
where ${\cal M}=(1+\beta)m_1+m_2$. Note $g_2(z)$ is now linear
and its sole zero $z=0$ is again not a image position.
Provided that $\zs\ne0$, the number of images for two perfectly aligned
lenses is therefore four \citep{wern08}.
All four images for an arbitrary source position $\zs$
are found to be $z=r{\rm e}^{{\rm i}\phi}$
where $\phi=\arg(\zs)$ -- i.e.\ $\zs=\lvert\zs\rvert{\rm e}^{{\rm i}\phi}$
-- and $r$ being the root of $r^4 - \lvert\zs\rvert r^3 - {\cal M} r^2
+ \beta m_1 \lvert\zs\rvert r + \beta m_1^2=0$.
Finally the solution for $\zs=0$ case, i.e.\
$r^4-{\cal M}r^2+\beta m_1^2=0$
corresponds the radii of two Einstein rings \citep{wern08}.


\section{The magnification of a source along the real axis}

Assuming $z_2\in\mathbf R$, we can evaluate the magnification
for the source along the real axis ($\zs\in\mathbf R$).
As noted, the quartic equation $g_4(z)=0$ yields four images on the real axis;
whereby i) two positive parity images and two negative parity images
if the source is outside caustics or
ii) three negative parity and one positive parity images
if the source is inside caustics.
On the other hand, the quadratic equation $g_2(z)=0$ for the source
inside caustics yields a pair of positive-parity off-axis images
of the equal magnification that are symmetric about the real axis. 
The real axis passes caustics through its cusp points.
As the source moves along the real axis and enters the caustic,
one positive-parity on-axis image crosses the critical curve and
turns into a negative-parity one whilst two positive-parity off-axis
images emerge from the same critical point passed by the on-axis image.
Upon the exit of the source from the caustic, the reverse process
(viz.\ merging of two positive-parity off-axis images
and one negative-parity on-axis image into one positive-parity on-axis
image on the critical point) takes place.

\subsection{A source inside caustics}

The image of positive parity on the real axis
is mostly very faint inside the {\it bright} caustic.
However, the situation changes inside the {\it faint} caustic.
In this case the off-axis magnification becomes $\mu \sim 0.1$ and  
the on-axis magnification of the positive parity becomes $\mu \geq 1$.  
Next, we further restrict ourselves to the case $M_1=M_2$ so that
$m_1 = 1 /(2-\beta)$ and $m_2 = (1-\beta) / (2-\beta)$.
However, we note here that we can partly expand our results
to unequal masses $M_1$ and $M_2$
because $\alpha = 1-\beta$ may be scaled like $\alpha' = \alpha M_2 / M_1$
as long as $ 0 < \alpha, \alpha' < 1 $ holds.
Applying the resultant method to the magnification for eliminating
$z$ and $\bar{z}$, we express the magnification as a function of $\zs$.
For the two positive-parity images resulting from $g_2(z)=0$, we find
\beq
\mu_1 = \mu_2 =  (1-\beta)^2 (\beta \zs -z_2)^2 z_2^2 \bigl[P_6(\zs)\bigr]^{-1}
\eeq
where $P_6(\zs) = \sum_{k=0}^6 c_k \zs^k$ is a sextic polynomial of $\zs$
whose coefficients are given by
\begin{subequations}\begin{align} \label{eq:c6}
c_6 &= -(2-\beta)^2 \beta^2 z_2^2  \\ \label{eq:c5}
c_5 &= 2 (2-\beta) \beta z_2 \bigl[ \beta (1-\beta) (1+ z_2^2)+ 2 z_2^2 \bigr] \\
c_4 &= -\beta^2 (1-\beta)^2 -2 \beta (2-\beta) (1-\beta) (1+3\beta) z_2^2
         - (2-\beta)^2 (1+4\beta+\beta^2) z_2^4 \\
c_3 &= 2 z_2 \bigl[2 \beta (1-\beta)^2 + 2 (4-\beta^2) (1-\beta) z_2^2+
         (2-\beta)^2 (1+\beta) z_2^4   \bigr]\\
c_2 &=  - z_2^2 \bigl[ 2 (1-\beta)^2 (2+\beta) + 
          2 (2-\beta) (1-\beta) (1+3\beta) z_2^2 + (2-\beta)^2 z_2^4  \bigr]\\
c_1 &= 2 (1-\beta) z_2^3 [2 (1 -\beta) + (2 - \beta) z_2^2] \\
c_0 &= -(1 - \beta)^2 z_2^4
\end{align}\end{subequations}
Let $\mu_3$ be the magnification of the positive-parity on-axis image and 
$\mu_4$, $\mu_5$ and $\mu_6$ the magnification of the negative-parity ones
which are located on the real axis.
Then using eq.~(\ref{eq:signed_mu}) and $\mu_1=\mu_2$,
we obtain an expression for the total magnification
\beq \label{mu_tot} 
\mu_{\rm tot} = \sum_{i=1}^6 \lvert\mu_i\rvert
=\sum_{i=1}^3 \mu_i - \sum_{i=4}^6 \mu_i
=2\sum_{i=1}^3 \mu_i - 1
= 4 \mu_1  + 2\mu_3 - 1,
\eeq
which is the key equation to understand the different
behaviour of the two caustics. Whilst inside the {\it bright} caustic
$\mu_1 \geq 1$ and therefore $\mu_{\rm tot} > 3$,
we have inside the {\it faint}
caustic $\mu_3 \geq 1$ and therefore $\mu_{\rm tot} > 1$. 

For the four on-axis images we can also derive a polynomial (of the 4th degree)
for the magnification, which is too long to be presented here.
However, to obtain the total magnification we only need
to estimate the magnification of the single positive-parity image, which 
turns out always to be very faint inside the {\it bright} caustic. 
For small $\beta$ we obtain
\beq
\mu_3 \approx { (1 - \beta)^4 \beta^6 \over 16 z_2^8 + 64 \beta z_2^6
(2- \zs z_2 + z_2^2)  + .....} \ll 1 
\eeq
Note that for $\beta \ll 1$ and $\beta \approx 1$ the numerator becomes
extremely small.
Since $\mu_3$ can be neglected we can derive
the minimum magnification from the equation above.
$\mu_1$ has a minimum  when
$2\beta P_6(\zs) = (\beta \zs - z_2) {P_6^\prime(\zs)}$
holds.
For the {\it faint} caustic $\mu_3$ takes the dominant part
and the images of negative parity becomes rather faint 
( $2 \mu_1 < \lvert \sum_{i=4}^6 \mu_i \rvert \ll 1$).
For the minimum magnification inside the {\it faint} caustic
we can write ${\rm d}\mu_{\rm tot}/{\rm d}\zs=
4({\rm d}\mu_1/{\rm d}\zs)+2({\rm d}\mu_3/{\rm d}\zs)=0$
by using eq. (\ref{mu_tot}).
Since $\mu_{\rm tot} \approx 2 \mu_3 - 1$ for this case
each derivative must vanish at the approximate same location $\zs$.
Therefore we can apply 
$2\beta P_6(\zs) = (\beta \zs - z_2) {P_6^\prime(\zs)}$
for the location of the 
minimum magnification inside the {\it faint} caustic as well.

Finally we note that the caustic, i.e. cusps intersects with the real axis
when $P_6 (\zs) = 0$. In this case we obtain infinite magnification
for a point source. The real solutions give the range where 6 images can occur.
The two caustic may merge if $P_6$ has a double solution or if
$P_6 (\zs) = 0$ and $P_6^\prime(\zs) = 0$ holds.
Using these two equations we obtain the condition when the two
caustics may touch and merge.
\begin{multline}
(2-\beta)^6 z_2^{12} +  2 (2-\beta)^5 (-3+7 \beta) z_2^{10}
+ (2-\beta)^4 (-15 -3\beta +70 \beta^2 -63\beta^3 +27 \beta^4) z_2^8 
 \\
-(2-\beta)^3(1-\beta) (8+65 \beta +207 \beta^2 - 513 \beta^3 + 297 \beta^4)
z_2^6 
- (2-\beta)^2 (1-\beta)^2 \beta (40 + 595 \beta -758 \beta^2 +27 \beta^3)
z_2^4 \\
+ 8 (2-\beta) (1-\beta)^3 \beta^2 (-39 +31 \beta) z_2^2
+16 (1-\beta)^4 \beta^3 = 0
\end{multline}
The change in the number of caustics and cusps for the simple binary lens
with shear was first studied by \cite{witt93}.

\subsection{The velocity of the {\it faint} caustic}

It is necessary to estimate the motion of the {\it faint} caustic
since the two point masses are usually not gravitationally bound to each
other for $\beta > 0$. Therefore we have to consider some projection
effects due to the motion of the second point mass $z_2 (t)$.
Assume that the second mass has a relative motion $v_2 = {\rm d}z_2 /{\rm d}t$.
To compute the velocity along the entire caustic one may follow 
\citet{kundic93}. This method allows for general deflectors to compute
the dependence of the velocity of each caustic point on the motion of the stars,
i.e.\ point masses. 
For a full analytical treatment this method turns out to be rather
complicated, however for a rough estimate we may inspect the polynomial
$P_6$ above. The solutions of $P_6(z_s)=0$ yield the positions of the
cusps along the real axis.
Using now eqs. (\ref{eq:c6}) and (\ref{eq:c5}) we can write
\beq
\sum_{i=1}^6 z_{{\rm cusp},i} = - {c_5 \over c_6}
= { 2 z_2 [2 + \beta (1-\beta) ( z_2^{-2} + 1 ) ] \over (2-\beta) \beta }
= 2 \left( { 1+\beta \over \beta } z_2
+ { 1-\beta \over 2-\beta } { 1 \over z_2} \right)
\eeq
where $z_{{\rm cusp},i}$ denotes 4 real solutions of the position of the cusps
on the real axis and two complex spurious solutions.
We assume here that we have at least two separate caustics. 
i.e. 4 cusps on the real axis (cf. Fig. \ref{fig:lc-0.8}).
Assuming now the second point mass moves along the real axis we may write
\beq
\sum_{i=1}^6 \frac{{\rm d}z_{{\rm cusp},i}}{{\rm d}t}
= 2 \left( { 1+\beta \over \beta }
- { 1-\beta \over 2-\beta } { 1 \over z_2^2} \right) v_2,
\eeq 
For small $\beta$ the first term dominates which gives
us the approximate  velocity of the {\it faint} caustic.
Since we have two cusp position we find $v_{\rm caustic} \approx v_2 / \beta $
and $z_{\rm cusp} \approx z_2 / \beta$ to the first order.
The other solutions remain small $\sim 1$. This is verified
numerically and by inspecting $c_0 / c_6$ which yields the
product of all six solutions.
This means for small $\beta$ the {\it faint} caustic may move
rather rapidly across the source plane so that the chance to observe
a complete caustic crossing is much smaller than that for moderate $\beta$.
Motions along the $y$-axis must be of the order of ${\rm d}z_2/{\rm d}t$ due to 
the rotation invariance of the caustics around the origin
for $z_2 = r_2 {\rm e}^{{\rm i}\varphi_2}$.



\label{lastpage}

\end{document}